\documentclass[superscriptaddress,prl,showpacs,twocolumn]{revtex4}
\usepackage{amssymb}
\usepackage{amsmath}
\usepackage{graphicx}
\usepackage{bm}

\input{tcilatex}

\begin{document}

\title{Topological Anderson Insulator}
\author{Jian Li}
\affiliation{Department of Physics and Center of Computational and Theoretical Physics,
The University of Hong Kong, Pokfulam Road, Hong Kong}
\author{Rui-Lin Chu}
\affiliation{Department of Physics and Center of Computational and Theoretical Physics,
The University of Hong Kong, Pokfulam Road, Hong Kong}
\author{J. K. Jain}
\affiliation{Department of Physics, 104 Davey Lab, Pennsylvania State University,
University Park, PA 16802, USA}
\author{Shun-Qing Shen}
\affiliation{Department of Physics and Center of Computational and Theoretical Physics,
The University of Hong Kong, Pokfulam Road, Hong Kong}

\begin{abstract}
Disorder plays an important role in two dimensions, and is responsible for
striking phenomena such as metal insulator transition and the integral and
fractional quantum Hall effects. In this paper, we investigate the role of
disorder in the context of the recently discovered topological insulator,
which possesses a pair of helical edge states with opposing spins moving in
opposite directions and exhibits the phenomenon of quantum spin Hall effect.
We predict an unexpected and nontrivial quantum phase termed "topological
Anderson insulator," 
which is obtained by introducing impurities in a two-dimensional metal; here
disorder not only causes metal insulator transition, as anticipated, but is
fundamentally responsible for creating extended edge states. We determine
the phase diagram of the topological Anderson insulator and outline its
experimental consequences.
\end{abstract}

\maketitle

Over the last thirty years, investigation of two dimensional systems has
produced a series of striking phenomena and states. A one-parameter scaling
theory for non-interacting electrons demonstrates that arbitrarily weak
random disorder drives the system into an insulating state, known as the
Anderson insulator \cite{Abrahams79PRL}. In the presence of strong spin
orbit coupling or interactions, a metallic state in two dimensions (2D)
becomes possible, and a metal-insulator transition occurs at a nonzero
critical value of disorder strength \cite{Hikami80,Ando89PRB,Kravchenko94PRB}%
. The application of a magnetic field, which breaks time reversal symmetry,
creates dissipationless edge states, resulting in the remarkable phenomenon
of the integral quantum Hall effect \cite{Klitzing80PRL}. Inter-electron
interaction produces the fractional quantum Hall effect \cite{Tsui82PRL},
characterized by topological concepts such as composite fermions, fractional
charge, and fractional statistics \cite{QHE}. Disorder plays a crucial role
in the establishment of the quantized Hall plateaus.

The quantum Hall state constitutes a paradigm for a topological state of
matter, the Hall conductance of which is insensitive to continuous changes
in the parameters and depends only on the number of edge states, which are
unidirectional because of the breaking of the time reversal symmetry due to
the magnetic field. Recently, an analogous effect was predicted in a time
reversal symmetric situation: it was shown that a class of insulators, such
as graphene with spin orbit coupling \cite{Kane05PRL} and an "inverted"
semiconductor HgTe/CdTe quantum well \cite{Bernevig06SCI}, possess the
topological property that they have a single pair of counter-propagating or
helical edge state, exhibiting the phenomenon of the quantum spin Hall
effect. This \textquotedblleft topological insulator" is distinguished from
an ordinary band insulator by a $Z_{2}$ topological invariant \cite%
{Kane05PRLb}, analogous to the Chern number classification of the quantum
Hall effect \cite{Thouless82PRL}. The edge states are believed to be
insensitive to weak (non-magnetic) impurity scattering \cite{Kane05PRLb} and
weak interaction \cite{Moore,Wu06PRL}. The prediction of non-zero
conductance in a band-insulating region of an "inverted" HgTe/CdTe quantum
well has been verified experimentally\cite{Konig07SCI}, although the origin
of the observed deviation from an exact quantization is not yet fully
understood. The topological insulator has also been generalized to three
dimensions \cite{Fu07PRL,Murakami07NJP,Hsieh08Nature}.

In view of its importance in 2D, it is natural to ask how disorder affects
the stability of the helical edge states in the topological insulator, which
has motivated our present study. As expected, we find that the physics of
topological insulator is unaffected by the presence of weak disorder but is
destroyed for large disorder. More surprisingly, however, our results show
that disorder can create a topological insulator for parameters where the
system was metallic in the absence of disorder, and also when the band
structure of the HgTe/CdTe quantum well is not inverted (i.e., the gap is
positive). We call this phase topological Anderson insulator (TAI) and
comment on the feasibility of its experimental observation.

The effective Hamiltonian for a clean bulk HgTe/CdTe quantum well is given
by \cite{Bernevig06SCI} 
\begin{equation}
\mathcal{H}(\bm{k})=\left( 
\begin{array}{cc}
h(\bm{k}) & 0 \\ 
0 & h^{\ast }(-\bm{k})%
\end{array}%
\right) ,  \label{eq:ham}
\end{equation}%
where $h(\bm{k})=\epsilon (k)+\bm{d}(\bm{k})\cdot \bm{\sigma}$, $\bm{k}$ $%
=(k_{x},k_{y})$ is the two-dimensional wave vector, $\bm{\sigma}%
=(\sigma_x,\sigma_y,\sigma_z)$ are Pauli matrices, and to the lowest order
in $k$, we have 
\begin{equation}
\bm{d}(\bm{k})=(Ak_{x},\,Ak_{y},\,M-Bk^{2});\;\epsilon (k)=C-Dk^{2}
\label{eq:dkeps}
\end{equation}%
with $A$, $B$, $C$ and $D$ being sample-specific parameters for which we
take realistic values from the experiment.\cite{Konig08ISPS} The lower
diagonal block $h^{\ast }(-\bm{k})$ is the time reversal counterpart of the
upper diagonal block. This four-band model describes the band mixture of the
s-type $\Gamma _{6}$ band and the p-type $\Gamma _{8}$ band near the $\Gamma 
$ point, in the basis of $|E1,m_{J}=+\frac{1}{2}\rangle $, $|H1,m_{J}=+\frac{%
3}{2}\rangle $ and $|E1,m_{J}=-\frac{1}{2}\rangle $ , $|H1,m_{J}=-\frac{3}{2}
\rangle $, where the $E1$ sub-band consists of linear combinations of $%
|\Gamma _{6},m_{J}=\pm \frac{1}{2}\rangle $ and $|\Gamma _{8},m_{J}=\pm 
\frac{1}{2}\rangle $ states, and the $H1$ sub-band consists of $|\Gamma
_{8},m_{J}=\pm \frac{3}{2}\rangle $ states. A band inversion results when $M$%
, the energy difference of the $E1$ and $H1$ states at the $\Gamma $ point,
changes its sign from positive to negative. Bernevig, Hughes and Zhang \cite%
{Bernevig06SCI} predicted that this sign change signifies a topological
quantum phase transition between a conventional insulating phase ($M>0$) and
a phase exhibiting the quantum spin Hall effect with a single pair of
helical edge states ($M<0$); experiments have confirmed some aspects of this
prediction.\cite{Konig07SCI}

\begin{figure}[tbp]
\centering \includegraphics[width=0.45\textwidth]{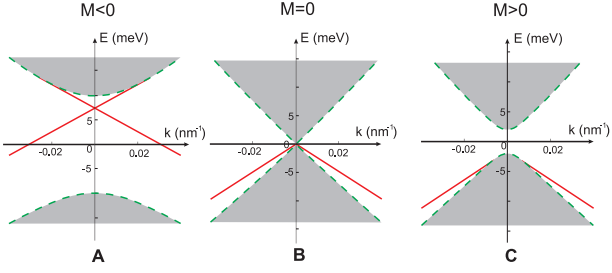}
\caption{Band structure of HgTe/CdTe quantum wells with finite width. (%
\textbf{A}) The {\textquotedblleft }inverted" band structure case with $%
M=-10 $ meV. Edge states (red lines) cross the bulk band gap and merge into
bulk states (gray area) at a maximum energy in the upper band. The green
dashed lines mark the boundary of bulk states. (\textbf{B}) The transition
point between an {\textquotedblleft }inverted" band structure and a {%
\textquotedblleft }normal" band structure with $M=0$ meV. (\textbf{C}) The {%
\textquotedblleft }normal" band structure with $M=2$ meV. In all figures,
the strip width $L_{y}$ is set to 100$\protect\mu $m. The sample-specific
parameters are fixed for all calculations in this paper to be $A=364.5$
meVnm, $B=-686$ meVnm$^{2}$, $C=0$, $D=-512$ meVnm$^{2}$. }
\label{fig:bands}
\end{figure}

In an infinite-length strip with open lateral boundary conditions, the
solution of the four-band model $\mathcal{H}\Psi =E\Psi $ is given by\cite%
{Zhou08xxx} 
\begin{eqnarray}
&&\Psi (k_{x},y)=\left( 
\begin{array}{c}
\bm{\psi}(k_{x},y) \\ 
\mathcal{T}\bm{\psi}(k_{x},y)%
\end{array}%
\right) ,  \notag  \label{eq:solution} \\
&&\bm{\psi}(k_{x},y)=\bm{\mu}_{+}e^{\alpha y}+\bm{\mu}_{-}e^{-\alpha y}+%
\bm{\nu}_{+}e^{\beta y}+\bm{\nu}_{-}e^{-\beta y}
\end{eqnarray}%
where $\mathcal{T}$ is the time-reversal operator, $\bm{\mu}_{\pm }$ and $%
\bm{\nu}_{\pm }$ are two-component $k_{x}$-dependent coefficients, and $%
\alpha $, $\beta $ are determined self-consistently by 
\begin{eqnarray}
&&\alpha ^{2}=k_{x}^{2}+F-\sqrt{F^{2}-(M^{2}-E^{2})/(B^{2}-D^{2})}
\label{eq:alpha} \\
&&\beta ^{2}=k_{x}^{2}+F+\sqrt{F^{2}-(M^{2}-E^{2})/(B^{2}-D^{2})}
\label{eq:beta} \\
&&E_{\alpha }^{2}\beta ^{2}+E_{\beta }^{2}\alpha ^{2}-\gamma E_{\alpha
}E_{\beta }\alpha \beta =k_{x}^{2}(E_{\alpha }-E_{\beta })^{2}.
\label{eq:secular}
\end{eqnarray}%
Here, we have $F=[A^{2}-2(MB+ED)]/2(B^{2}-D^{2})$, $E_{\alpha
}=E-M+(B+D)(k_{x}^{2}-\alpha ^{2})$, $E_{\beta }=E-M+(B+D)(k_{x}^{2}-\beta
^{2})$, $\gamma =\tanh {\frac{\alpha L_{y}}{2}}/\tanh {\frac{\beta L_{y}}{2}}%
+\tanh {\frac{\beta L_{y}}{2}}/\tanh {\frac{\alpha L_{y}}{2}}$, and $L_{y}$
is the width of the strip. This solution naturally contains both helical
edge states ($\alpha ^{2}<0$) and bulk states ($\alpha ^{2}>0$), which are
shown in Fig. \ref{fig:bands} for three cases $M<0$, $M=0$, and $M>0$. The
edge states (red lines in Fig. \ref{fig:bands}) are seen beyond the bulk gap
for all cases, up to an $M$ dependent maximum energy. When $M<0$, the edge
states cross the bulk gap producing a topological insulator. At $M=0$ the
edge states exist only in conjunction with the lower band, terminating at
the Dirac point. For positive $M$ there are no edge states in the gap,
producing a conventional insulator.\cite{Bernevig06SCI}

\begin{figure}[tbph]
\label{fig:pds}\centering \includegraphics[width=0.45\textwidth]{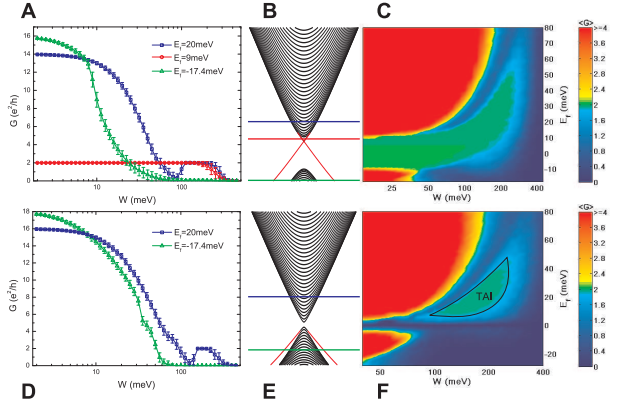}
\caption{Conductance of disordered strips of HgTe/CdTe quantum wells. The
upper panels (A to C) show results for an quantum well {\textquotedblleft }%
inverted" with $M=-10$ meV, and the lower panels (D to F) for a {%
\textquotedblleft }normal" quantum well with $M=1$ meV. (\textbf{A}) The
conductance $G$ as a function of disorder strength $W$ at three values of
Fermi energy. The error bars show standard deviation of the conductance for $%
1000$ samples. (\textbf{B}) Band structure calculated with the tight-binding
model. Its vertical scale (energy) is same as in (C) and the horizontal
lines correspond to the values of Fermi energy considered in (A). (\textbf{C}%
), Phase diagram showing the conductance $G$ as a function of both disorder
strength $W$ and Fermi energy $E_{f}$. The panels (\textbf{D}), (\textbf{E}%
), and (\textbf{F}) are same as (A), (B), and (C), but for $M>0$. The TAI
phase regime is labeled. In all figures, the strip width $L_{y}$ is set to $%
500$ nm; the length $L_{x}$ is $5000$ nm in (A) and (D), and $2000$ nm in
(C) and (F).}
\end{figure}

We next study transport as a function of disorder, with the Fermi energy
varying through all regions of the band structure. For this purpose, we use
a tight-binding lattice model which produces the above Hamiltonian as its
continuum limit\cite{Bernevig06SCI}, and following a common practice in the
study of Anderson localization, introduce disorder through random on-site
energy with a uniform distribution within $[-W/2,W/2]$. We calculate the
conductance of disordered strips of width $L_{y}$ and length $L_{x}$ using
the Landauer-B\"{u}ttiker formalism.\cite{Landauer,Buttiker} The conductance 
$G$ as a function of disorder strength $W$ is plotted in Fig. \ref{fig:pds}
for several values of Fermi energy belonging to different band regions for $%
M<0$ and $M>0$. The topological nature of the system is revealed by the
quantization of conductance at $2e^{2}/h$. The following observations can be
made.

The calculated behavior conforms to the qualitative expectation for certain
situations. For Fermi level in the lower band, for both $M<0$ and $M>0$, an
ordinary Anderson insulator results when the clean limit metal is disordered
(green lines in Fig. \ref{fig:pds}A and \ref{fig:pds}D). The conductance in
this case decays to zero at disorder strength around $100$ meV, which is
about two times the leading hopping energy $t_{max}=(|B|+|D|)/a^{2}\approx
48 $ meV, and much larger than the clean-limit bulk band gap $E_{g}=2|M|=20$
meV. Here $a=5$ nm is the lattice spacing of the tight-binding model. This
energy scale is several times smaller than the critical disorder strength $%
W_{c}\approx 350$ meV for the metal-insulator transition in such a system in
2D, as extracted from one-parameter scaling calculations \cite{note1}. The
topological insulator (red line in Fig. \ref{fig:pds}a) is robust, and
requires a strong disorder before it eventually yields to a localized state.
This is expected as a result of the absence of backscattering in a
topological insulator when time-reversal symmetry is preserved\cite%
{Kane05PRLb}. 

The most surprising aspect revealed by our calculations is the appearance of
anomalous conductance plateaus at large disorder for situations when the
clean limit system is a metal without preexisting edge states. See, for
example, the blue lines in Fig. \ref{fig:pds}A ($M<0$) and Fig. \ref{fig:pds}%
D ($M>0$). The anomalous plateau is formed after the usual metal-insulator
transition in such a system. The conductance fluctuations (the error bar in
Fig. \ref{fig:pds}A and \ref{fig:pds}D) are vanishingly small on the
plateaus; at the same time the Fano factor drops to nearly zero indicating
the onset of dissipationless transport in this system,\cite{Buttiker90} even
though the disorder strength in this scenario can be as large as several
hundred meV. This state is termed topological Anderson insulator. The
quantized conductance cannot be attributed to the relative robustness of
edge states against disorder, because it occurs for cases in which no edge
states exist in the clean limit! The irrelevance of the clean-limit edge
states to this physics is further evidenced from the fact that no anomalous
disorder-induced plateaus are seen for the clean limit metal for which bulk
and edge states coexist; those exhibit a transition into an ordinary
Anderson insulator.

The nature of TAI is further clarified by the phase diagrams shown in Fig. %
\ref{fig:pds}C for $M<0$ and in Fig. \ref{fig:pds}F for $M>0$. For $M<0$,
the quantized conductance region (green area) of the TAI phase in the upper
band is connected continuously with the quantized conductance area of the
topological insulator phase of the clean-limit. One cannot distinguish
between these two phases by the conductance value. When $M>0$, however, the
anomalous conductance plateau occurs in the highlighted green island labeled
TAI, surrounded by an ordinary Anderson insulator. No plateau is seen for
energies in the gap, where a trivial insulator is expected. The topology of
the TAI phase as well as the absence of preexisting edge states in the clean
limit demonstrate that the TAI owes its existence fundamentally to disorder.

\begin{figure}[tbp]
\centering \includegraphics[width=0.45\textwidth]{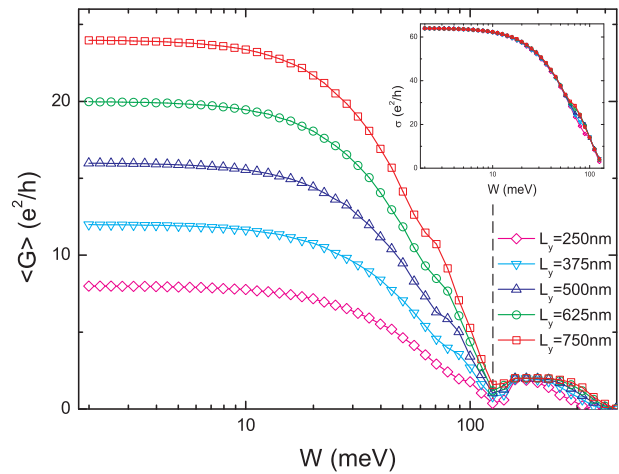}
\caption{Width-dependence of the conductance in disordered strips of
HgTe/CdTe quantum wells. The conductance as a function of disorder strength
is plotted for several values of strip width, with fixed length $L_{x}=2000$
nm. In the inset, the conductance traces prior to the TAI phase
(left-hand-side of the dashed line) are scaled with the width of the strips,
and are presented as the conductivity defined as $\protect\sigma %
=GL_{x}/L_{y}$. The formation of the edge states is indicated by the
presence of conductance quantization $2e^{2}/h.$ In this figure, we take $%
M=2 $ meV, and the Fermi energy is set to $E_{f}=20$ meV.}
\label{fig:size}
\end{figure}

The dissipationless character suggests existence of ballistic edge states in
the TAI phase. To gain insight into this issue, we investigate how the
conductance scales with the width of the strip. Fig. \ref{fig:size} shows
the calculated conductances of a strip as a function of its width $L_{y}$.
In the region before the TAI phase is reached, the scaled conductance $%
GL_{x}/L_{y}$, or conductivity, is width independent, as shown in the inset
of Fig. \ref{fig:size}, which implies bulk transport. Within the TAI phase,
absence of such scaling indicates a total suppression of the bulk
conduction, thus confirming presence of conducting edge states in an
otherwise localized system.

\begin{figure}[tbp]
\centering \includegraphics[width=0.45\textwidth]{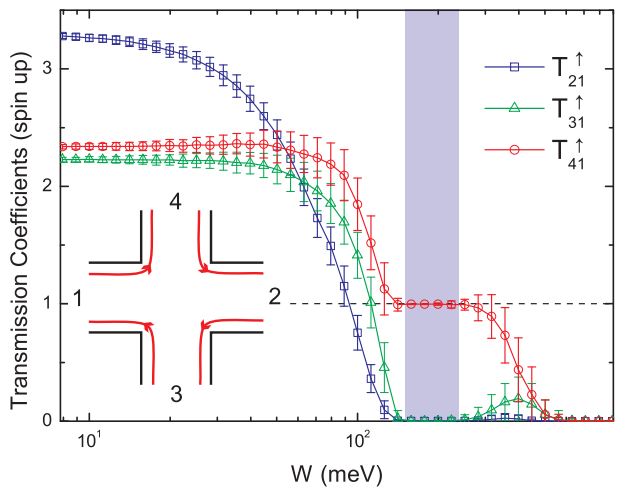}
\caption{The quantum spin Hall effect in a TAI-based Hall bar. Three
independent spin-resolved transmission coefficients, $T_{21}^{\uparrow }$, $%
T_{31}^{\uparrow }$ and $T_{41}^{\uparrow }$, are plotted as functions of
disorder strength $W$, where $T_{pq}^{\uparrow }$ stands for the
transmission probability of spin-up electrons from lead $q$ to lead $p$
(lead indices are shown in the inset). Standard deviations of the
transmission coefficients for 1000 samples are shown as the error bars. In
the shadowed range of disorder strength, all bulk states are localized and
only chiral edge states exist, which is schematically shown in the inset
(for spin-up component only). The width of leads is $500$ nm and the other
parameters are set at $M=1$ meV and $E_{f}=20$ meV.}
\label{fig:crossbar}
\end{figure}

We further examine the picture of edge-state transport in the TAI phase in a
four-terminal cross-bar setup by calculating the spin resolved transmission
coefficients $T_{pq}^{s}$ ($s=\uparrow ,\downarrow $) between each ordered
pair of leads $p$ and $q$ ($=1,2,3,4$). Time-reversal symmetry guarantees
that $T_{pq}^{\uparrow }=T_{qp}^{\downarrow }$, so it suffices to discuss
only one spin component. Three independent coefficients, $T_{21}^{\uparrow }$%
, $T_{31}^{\uparrow }$ and $T_{41}^{\uparrow }$, are shown in Fig. \ref%
{fig:crossbar} as a function of the disorder strength inside the cross
region. The shadowed area marks the TAI phase, where $\langle
T_{41}^{\uparrow }\rangle =1$, $\langle T_{21}^{\uparrow }\rangle =\langle
T_{31}^{\uparrow }\rangle =0$, and all transmission coefficients exhibit
vanishingly small fluctuations. From symmetry, it follows that 
\begin{equation}
\langle T_{41}^{\uparrow }\rangle =\langle T_{24}^{\uparrow }\rangle
=\langle T_{32}^{\uparrow }\rangle =\langle T_{13}^{\uparrow }\rangle
\rightarrow 1,
\end{equation}%
and all other coefficients are vanishing small. These facts are easy to
understand from the presence of a chiral edge state for the spin up block in
the TAI phase. Two consequences of this chiral edge state transport are a
vanishing diagonal conductance $G_{xx}^{\uparrow }=(T_{21}^{\uparrow
}-T_{12}^{\uparrow })e^{2}/h=0$ and a quantized Hall conductance $%
G_{xy}^{\uparrow }=(T_{41}^{\uparrow }-T_{42}^{\uparrow })e^{2}/h=e^{2}/h$,
analogous to Haldane's model for the integer quantum Hall effect with parity
anomaly.\cite{Haldane88PRL} The quantized Hall conductance $G_{xy}^{\uparrow
}$ reveals that the topologically invariant Chern number of this state is
equal to one.\cite{Thouless82PRL,Niu85PRB,Lee08PRL} Since the Hamiltonian
for spin down sector $h^{\ast }(-\bm{k})$ is the time reversal counterpart
of $h(\bm{k})$ for the spin up sector, we have the relations $%
T_{pq}^{\uparrow }=T_{qp}^{\downarrow }.$ The absence of Hall current in a
time reversal invariant system implies $G_{xy}^{\downarrow
}=-G_{xy}^{\uparrow }=-e^{2}/h$. Thus, the chiral edge state in the spin
down sector moves in the opposite direction as the edge state in the spin up
sector. As a result, the total longitudinal conductance and Hall conductance
both vanish as in an ordinary insulator, but the dimensionless spin Hall
conductance $G_{xy}^{s}=(G_{xy}^{\uparrow }-G_{xy}^{\downarrow
})/(e^{2}/h)=2 $, resulting in quantum spin Hall effect.\cite{Kane05PRL}

Our work thus predicts quantized conductance in the presence of strong
disorder, even when the band structure has positive gap (i.e. no inversion)
and the clean limit system is an ordinary metal. We believe that HgTe/CdTe
quantum wells, which have been used to investigate topological insulators in
the clean limit, are a promising candidate also for an experimental
determination of the phase diagram of the topological insulator as a
function of disorder and doping, because they enable a control of various
parameters through variations in the quantum well thickness and gate voltage.%
\cite{Becker00PRB,Zhang01PRB} The effect of disorder in three dimensional
topological insulators should also prove interesting.

The authors would like to thank F. C. Zhang for discussions. This work was
supported by the Research Grant Council of Hong Kong under Grant No. HKU
7037/08P.


\begin{thebibliography}{99}
\bibitem{note1} The one-parameter scaling calculations indicate that the
metal-insulator transition occurs in the 2D system of this material where
electron-electron interaction is absent, and the term $d_{x}\sigma
_{x}+d_{y}\sigma _{y}$ in Eq. \eqref{eq:ham} plays the role of spin-orbit
interaction like the Rashba coupling.
\end{thebibliography}
\end{document}